# Artificial intelligence in government: Concepts, standards, and a unified framework

Vincent J. Straub[1*], Deborah Morgan[1,2], Jonathan Bright[1*], and Helen Margetts[1,3]

[1] Alan Turing Institute
[2] Accountable, Responsible and Transparent AI CDT,
Department of Computer Science, University of Bath
[3] Oxford Internet Institute, University of Oxford, Oxford, UK

[*] Corresponding authors: vincejstraub@gmail.com, jbright@turing.ac.uk

**Abstract.** Recent advances in artificial intelligence (AI), especially in generative language modelling, hold the promise of transforming government. Given the advanced capabilities of new AI systems, it is critical that these are embedded using standard operational procedures, clear epistemic criteria, and behave in alignment with the normative expectations of society. Scholars in multiple domains have subsequently begun to conceptualize the different forms that AI applications may take, highlighting both their potential benefits and pitfalls. However, the literature remains fragmented, with researchers in social science disciplines like public administration and political science, and the fast-moving fields of AI, ML, and robotics, all developing concepts in relative isolation. Although there are calls to formalize the emerging study of AI in government, a balanced account that captures the full depth of theoretical perspectives needed to understand the consequences of embedding AI into a public sector context is lacking. Here, we unify efforts across social and technical disciplines by first conducting an integrative literature review to identify and cluster 69 key terms that frequently co-occur in the multidisciplinary study of AI. We then build on the results of this bibliometric analysis to propose three new multifaceted concepts for understanding and analysing AI-based systems for government (AI-GOV) in a more unified way: (1) *operational fitness*, (2) *epistemic alignment*, and (3) *normative divergence*. Finally, we put these concepts to work by using them as dimensions in a conceptual typology of AI-GOV and connecting each with emerging AI technical measurement standards to encourage operationalization, foster cross-disciplinary dialogue, and stimulate debate among those aiming to rethink government with AI.

## 1 Introduction

The adoption of new technology is seen as one of the main routes to transform government. But although the public sector has frequently been at the forefront of technological progress, innovation has often backfired (Margetts, 1999). This has become especially salient in an increasingly data-intensive, digital world as administrations still largely designed for very different times fail to successfully provide basic online services and regularly do not deliver on flagship policies reliant on information and communication technologies (ICTs). The UK government's NHS COVID-19 app announced in 2020 to monitor the spread of the SARS-CoV-2 virus, for instance, was abandoned after a few months in favour of a superior system developed by tech companies. If governments are to make effective use of new technologies while maintaining citizen trust, scholars and partitioners need to develop a richer theoretical understanding of these tools and systems.

The need for new theories and conceptual frameworks is especially pertinent for understanding the latest technology to hold the promise of fundamentally transforming government: artificial intelligence (AI), defined by the EU High-Level Expert Group as "systems that display intelligent behaviour by analysing their environment and taking actions—with some degree of autonomy—to achieve specific goals" (Ala-Pietilä & Smuha, 2021). Following a decade of accelerated progress, predominantly in the paradigm of machine learning (ML) (Domingos, 2012), contemporary AI systems, especially generative AI applications developed using deep learning (Bengio et al., 2021) like large language models (LLMs) (e.g., BERT, DALL-E, and GPT-4) (Bommasani et al., 2022), arguably represent a step-change from the era of simpler ICTs (Margetts, 2022). Thanks to the availability of computing resources coupled with increasing dataset sizes and advances in modelling techniques, LLMs and other state-of-the-art systems possess advanced capabilities (e.g., in language, speech, vision, robotics, human interaction) that allow them to execute an increasing range of government-related occupational tasks (Brynjolfsson & Mitchell, 2017; Eloundou et al., 2023). As a result, scholars have increasingly framed AI as a general-purpose technology, one that will have a protracted impact by identifying solutions in one sector that are generalizable to problems in other sectors (Brynjolfsson et al., 2018; Crafts, 2021).

AI applications have already been adopted by various public agencies, organisations that implement government policies and may contribute to their development, in the hope of making government services more effective and responsive. In the context of the United States, one comprehensive study found that 45 percent of federal agencies had experimented with AI by 2020 (Engstrom et al., 2020). Key tasks that governments have been able to trial AI for to improve service delivery include detection, prediction, and simulation (Margetts & Dorobantu, 2019). Notable applications range from facial-recognition in policing (Zilka et al., 2022) and recidivism prediction in criminal justice (Kleinberg et al., 2018) to the use of virtual agents in process automation (Ojo et al., 2019) and forecasting future needs in social services (Bright et al., 2019). Yet, as more public agencies begin to adopt these applications, it is critical that they are embedded using standard operational procedures, clear epistemic criteria, and are in alignment with the normative concerns of society. As pointed out by Köbis et al. (2022), this is because AI systems, thanks to their learning capabilities, can—in contrast to earlier generations of ICTs—act autonomously and even self-modify their behaviour (Rahwan et al., 2019). The adoption of these advanced tools is in turn riddled with "AI tensions" (Madan & Ashok, 2022): technical and social challenges relating to fairness, transparency, and explainability, among others. Indeed, a growing body of evidence is revealing the perils and pitfalls of embracing AI systems, especially in terms of their complex algorithmic design (Adadi & Berrada, 2018) and biased input data (Buolamwini & Gebru, 2018), alongside their variable performance (Dressel & Farid, 2018) and functionality failures (Raji et al., 2022). Case-study work has shed further led light on the issue, revealing that the introduction of AI can bring about profound and at times disruptive shifts in both organizational structure and bureaucratic agency (Giest & Klievink, 2022), both at a national and subnational level (Vogl et al., 2020).

In response to the substantive harms already caused by existing AI applications in high-stakes decisions (McGregor, 2020; Rudin, 2019), there have been urgent calls for each stage in the AI lifecycle and value chain—from education and research to design and adoption—to not exist in technological isolation without greater input from scholars of human behaviour, ethics, culture, and law (Grosz et al., 2019; Veale, 2020; Widder & Nafus, 2022; Straub et al., 2023). As a result, many institutions, researchers, and thought leaders have advanced the notion that AI systems should be 'human-centered' (Shneiderman, 2020). Yet, aside from highlighting the need for AI systems to be reliable, safe, and trustworthy for humans, there is still no consensus definition for what this term means. Taken together, the emerging study of AI in government is hence characterised by a huge amount of interest from a very wide variety of scholarly disciplines (as we will elaborate further below). This variety has, however, gone hand in hand with some fragmentation in theory and terminology, with researchers in public administration, political science, policy studies and the fast-moving fields of AI, ML and robotics developing concepts in relative isolation. This is substantiated by the fact that technical research regularly fails to take non-technical viewpoints or the wider deployment context into account: a recent survey of ML research found that only 15% papers justify how their project connects to a societal need and just 1% discuss negative potential (Birhane et al., 2022). Moreover, because different aspects of AI in government have often been studied in different disciplines, many of the concepts scholars use overlap in meaning or are used indiscriminately, such as (to take but a few examples) the concepts of 'comprehensibility', 'intelligibility', and 'understandability' (Markus et al., 2021). At the same time, the increasing number of typologies and taxonomies now being developed to understand the use of AI in government often do not acknowledge let alone build on each other, perpetuating the "fragmented state of knowledge" (Wirtz et al., 2019). While researchers have repeatedly highlighted this problem (Kankanhalli et al., 2019; Sloane & Moss, 2019; Sousa et al., 2019; Zuiderwijk et al., 2021), a balanced account of human-centered AI for government (AI-GOV), one that offers a shared conceptual language and captures the full depth of disciplinary perspectives, is critically needed to understand the consequences of embedding machine intelligence into government.

In this article, we seek to unify analytical efforts across social and technical disciplines by first conducting an integrative literature review to identify and cluster key terms that frequently co-occur in the multidisciplinary study of AI. We then build on the results of this bibliometric analysis by proposing three new multifaceted concepts for understanding and analysing AI-GOV in a more unified way: (1) *operational fitness*, (2) *epistemic alignment*, and (3) *normative divergence*. Finally, we use these concepts as dimensions to construct a conceptual typology of different types of AI-GOV and link each to emerging technical AI measurement standards, recommended metrics and methods for the measurement and evaluation of AI technologies (AIME Planning Team, 2021; British Standards Institution, 2022), in an effort to encourage the operationalization of concepts, foster cross-disciplinary dialogue and stimulate debate among those aiming to reshape public administration with AI. To motivate these contributions, we first provide an overview of the different ways the study of AI in government has so far been approached. We then describe our literature search method in Section 3, before presenting the results of our concept co-occurrence analysis in Section 4. In Section 5, we outline three new multi-faceted concepts and present a typology of AI-GOV, drawing links between each dimension and existing AI measurement standards alongside discussing the practical implications of our typology.

We conclude by highlighting future research avenues that we anticipate will become more pressing in future.

## 2 Background and related work

To understand the need for a more exhaustive account of AI-GOV, we must take a closer look at the term 'AI' and consider which terms we should use to frame its use in a public sector context. To do so we first define AI systems, consider how the problems and risks AI poses are fundamentally different to older ICTs and review how AI in government has so far been defined in related work, focusing on which concepts have featured in past attempts to offer a unified perspective.

While a ('general') intelligent machine remains a purely hypothetical concept and a universally accepted definition of AI remains elusive (Bubeck et al., 2023; Wang, 2019), current AI systems tend to either be software-focused (e.g., recidivism prediction instruments) or tied to certain hardware devices (e.g., autonomous drones) (Ala-Pietilä et al., 2019). A further distinction can arguably be made between systems based on the degree of autonomy and self-modification capabilities they possess. Autonomous AI agents, such as self-driving cars, present distinct social dilemmas as they will sometimes have to choose between two undesirable real-world outcomes—running over pedestrians or sacrificing their passenger to save them, for example. Figuring out how to build ethical autonomous machines consequently remains one of the thorniest challenges in AI research (see Bonnefon et al., 2016 for a discussion). Meanwhile, machines that can 'evolve' by modifying their underlying architecture without learning from new training data or being explicitly programmed to do arguably represent a separate risk class altogether. This is because such systems can make explainability impossible if humans have to reverse-engineer algorithms to understand a machine's behaviour—which may itself not be possible if it involves violating the terms of service of the company who built the system. These risks become even more complex in hybrid systems composed of many machines and humans; some researchers have consequently called for a new field of 'machine behaviour' to understand AI systems that possess these attributes (Rahwan et al., 2019). Yet, despite these risks, the 'intelligent' capabilities of state-of-the-art applications like OpenAI's large multimodal model GPT-4 (OpenAI, 2023) are currently still fundamentally built around correlation and association (Bender et al., 2021). Nevertheless, alongside helping to reignite the field of AI, they are encouraging researchers to build bridges across different research fields. This emergent integration of data-intensive ML with more traditional, symbolic approaches coupled with human feedback (Salmen, 2022) is in turn providing increased support for AI's promise as a horizontal technology (Leslie, 2019).

Since we are pursuing a unified account of AI in government, we follow the definitional path trodden by studies on the potential of AI (Gil & Selman, 2019). In the rest of this article, we in turn use the term 'AI system' broadly to refer to a data-driven, machine-based system that can complete—with some degree of autonomy and learning ability—a specific cognitive task (or set of tasks), using ML, agent computing, or related techniques. Unsurprisingly, many other terms are also used in discussions of AI in government; 'algorithmic system' and 'automated decision system' feature prominently in policy and legal articles (Richardson, 2022), for instance. We adopt AI system, as this arguably better distinguishes contemporary systems from earlier generations of computer technology in terms of their unique, data-powered nature and advanced analytic capabilities (Pencheva et al., 2020). Adopting this definition also enables us to conduct a conceptual analysis that is not restricted only to existing ML-based applications, although much of the scholarship we cover focuses on this. Moreover, we further employ 'AI-GOV' to emphasize that all government AI systems operate within social and institutional contexts. This aligns our argument with socio-technical accounts of technological systems not as isolated computational components but entities integrated into larger structures that encompass infrastructures, human operators, and norms (Selbst et al., 2019). Put differently, information processing—from the collection of input data to the final prediction or analytical output—involves several interconnected steps executed by both humans and machines (Chen & Golan, 2016). When viewed in this way, as a type of 'human-machine network' (Tsvetkova et al., 2017), AI-GOV can be considered a function of the fit between a system (including both human and artificial agents) and its environment (task and institutional context) (Rahwan et al., 2019). Notably, this conception chimes with the original study of administrative behaviour as set out by Simon (1947, 1955).

Moving beyond the problem of defining AI, a broader issue brought to light in past literature is the question of deciding which concepts and terms are relevant for characterizing AI-GOV. Many conceptual pieces in public administration have already drawn attention to the need to consider the interaction between AI systems and the characteristics of their environment, such as the organizational setting and task complexity (Bullock et al., 2020; Bullock, 2019; van Noordt & Misuraca, 2022). At the same time, various speculative pieces have long argued that the likelihood of AI technologies either displacing or augmenting the abilities of government actors critically depends on how we design and conceive them (Barth and Arnold 1999; Brynjolfsson, 2022). Perhaps most urgently, there is also a growing awareness that AI research and development, more so than older ICTs, requires scholars to take non-technical viewpoints and the wider deployment context into account. This is arguably because of the concrete negative consequences that AI systems currently in use are already having on actual people, often in marginalized communities, as documented in research in the area of algorithmic injustice (see Marjanovic et al., 2022 for an overview). In the case of LLMs, these ethical and social risks include information hazards, misinformation harms, hate speech and exclusion to name but a few (see Weidinger et al., 2022 for a discussion). Despite these acute problems, interdisciplinary scholarship informed by the latest wave of AI in government is still nascent, and scholars agree that theory still needs to add real value to the work of practitioners (Birhane, 2021). A plethora of conceptual terms and the lack of integration, especially between social science and the technical fields of AI (Sloane & Moss, 2019), means governments can struggle to know which ideas, principles, and values to prioritize in developing and governing AI.

While we are by no means the first to seek a unified account of AI in government, prior attempts to conceptualize AI-GOV have arguably tended to default to unidimensional thinking. That is, they have relied heavily on a single viewpoint or concept—an abstract noun denoting both a quality or an idea that offers a point of view for understanding some aspect of experience (e.g., bias), and, relatedly, a mental image that can be operationalized (e.g., measurement bias)—to discuss AI. Perhaps most prominent, especially in studies that build on earlier literature on large-scale, so-called 'big data' (Mergel et al., 2016) and the emergence of data science (Provost & Fawcett, 2013), is simply the focus on the advanced analytic capabilities of core AI applications including computer vision, robotics, and natural language processing (Norvig & Russell, 2020). Terms that consequently feature heavily in such discussions on the potential of AI in government are 'automation' or 'efficiency'. Other prominent examples include 'discretion', in a public administration context (Bullock, 2019), or 'explainability', in more recent ML research (de Bruijn et al., 2022). A number of these studies are advanced in scope, bridging disciplines or decomposing each concept into subcomponents. For example, in considering the 'riskiness' of public sector predictive tools, Oswald (2018) takes into account both legal perspectives and the practicalities of the sociotechnical context. However, many conform to an approach epitomized by Mehr (2017): they use the term 'AI' strictly in a computational sense, foregrounding the data-driven decision-making capacity of AI systems, and distinguishing different systems predominantly in terms of the domain or "government problem" (ibid) they address (e.g., resource allocation) (Engstrom et al., 2020; Misuraca et al., 2020; Samoili et al., 2021). Yet, this line of reasoning leaves out many important ethical, legal, and institutional factors that characterize the reality of AI-GOV in practice.

More structured, theoretically-informed analyses often still resort to simplified categories to distinguish AI-GOV, such as 'automation-friendly' or '[decision] support' (Katzenbach & Ulbricht, 2019). Sætra (2021), for example, combines the concepts 'human-in the-loop' (Green & Chen, 2019), 'autonomy', 'transparency', and 'governance' to devise a typology of five types of 'political AI systems'. Yet, they make no mention of 'bias' or 'fairness'—concepts that have arguably become equally as important in discussions of government AI (Mehrabi et al., 2022). Beyond political science and public administration, similar criticisms can be levelled at attempts to theorize about AI systems from the outside in by applying social science concepts to computer science frameworks (Ballester, 2021). Dellermann et al. (2021), for instance, develop a taxonomy of human-AI systems to investigate the notion of 'hybrid intelligence'. Their framework does not, however, differentiate between systems based on 'trust and 'safety', two of the most salient concepts to emerge in recent policy debates on AI (Glikson & Woolley, 2020)—and ones that arguably needs to be part of any attempt to conceptualize the future of human-AI collaboration (Brundage et al., 2020). Perhaps the most exhaustive classifications of AI systems to date have been proposed by organisations like the National Institute for Standards and Technology (NIST, 2022), which has developed an AI Risk Management Framework that explicitly incorporates socio-technical attributes. Specifically, the framework is comprised of four 'core functions' (Govern, Map, Measure and Manage) to help organizations address the risks of AI systems, which includes performance considerations like reliability, as well as so-called 'trustworthiness characteristics', such as accountability (see Figure 4 in ibid.). This is similar to the OECD (2022) framework for the classification of AI systems developed to evaluate AI from a user-friendly policy perspective. The 5 key dimensions of this framework 'People & Planet', 'Economic context', 'Data & Input', 'AI Model', and 'Task & Output' are linked to different stages of the AI system's life cycle, and 3 explicitly include what can broadly be considered social aspects (i.e., the dimensions People & Planet, Data & Input, and AI Model consider whether a system's outputs impact fundamental human rights like privacy, process personal data, and are explainable).

While the study of AI in government is constantly evolving, prior reviews have found that the field as a whole remains fragmented, as there is a distinct lack of crossover between different literatures and lack of theoretical foundations (Kankanhalli et al., 2019; Wirtz et al., 2019). In the existing academic attempts to offer a formal account of human-machine systems and AI in government reviewed above, we find a similar pattern. That is, the majority of studies exhibit an integrative deficiency: they foreground specific viewpoints and concepts at the expense of unity. Much work in the social and policy sciences ultimately stresses the ethical and governance challenges at stake in the adoption of AI; while research in computer science tends to highlight the computational and operational aspects that need to be considered. Even in the case of the more exhaustive frameworks developed by NIST and the OECD, their wide remit, context-neutral and policy-friendly nature means they arguably achieve disciplinary breadth at the expense of interdisciplinary depth. That is, the broad dimensions they propose treat the technical aspects of an AI system, such as whether a model's outputs are explainable, as separate from the social implications, for instance, whether users of the system possess the competencies to understand, interact with and can contest its outputs. However, while conceptually convenient, this assumption arguably rests on a false dichotomy, as it perpetuates the view that AI systems themselves are ultimately isolated computational components rather than embedded socio-technical entities (as we conceive them above). While this is clearly useful in the short run for defining the material scope of AI systems (see Mökander et al., 2022), it contributes to the lack of integration between fields in the long run by distinguishing between concepts that should be studied from a social or a technical perspective. As pointed out by Burr and Leslie (2022) in discussing scholarship on AI ethics, addressing this shortcoming and uniting the field requires sustained interdisciplinary effort and a richer consideration of the socio-technical nature of AI-GOV (Straub et al., 2023). Taken together, the main contribution of our study is that we seek to identify and understand the myriad different concepts used in the multidisciplinary scholarship on AI in government in order to then propose three new, multifaceted concepts that better capture and integrate the complexities of the sociotechnical phenomena of AI-GOV, thereby encouraging deeper cross-disciplinary dialogue down the line.

# 3 Materials and methods

To capture the depth of analytical perspectives needed to understand AI-GOV, we advance a bottom-up approach and first map existing scholarship. Specifically, we conducted an integrative literature review search following the methodology proposed by Torraco (2016) and used the bibliometric method of keyword co-occurrence analysis (Segev, 2021) to identify concepts that have frequently featured in technical, social scientific, and humanistic literature concerned with AI-GOV. In mapping key concepts, we are not specifying which ones should be studied, rather, our purpose is to detect, and ultimately integrate, dominant conceptual trends in the contemporary study of AI in government. In Section 3.1. below we first describe our literature search and data collection procedure, which is summarized graphically in Figure 1. In Section 3.2, we outline the method we used to extract concepts from articles combining automatic keyword extraction and manual review.

## 3.1 Data collection

To identify key concepts in current scholarship on AI in government, a bibliometric approach was employed, which involved retrieving highly cited articles and then extracting relevant concepts from these articles. Specifically, we adopted a three-step study design comprising (i) search term selection, (ii) corpus construction and (iii) concept extraction, illustrated in Figure. 1A. Given the main aim and contribution of our paper is to offer a new conceptual framework, informed by current theorizing, that can foster cross-disciplinary exchange, our literature review search procedure is purposively not comprehensive, rather we restrict our search to influential (i.e., highly cited) articles. Searching was conducted on 11 March 2023, and was updated on 11 May 2023 using the Google Scholar (GS) (Halevi et al., 2017) bibliographical database, which provides citation metrics and indexes literature in the social sciences, humanities and technical fields like computer science. Following integrative literature review conventions (Torraco, 2016), the search strategy was based on first selecting a small number of highly cited 'seed articles' (n=7; see Fig. 1) on the topic of AI and government published within the last 5 years in the high-impact, top-tier field outlets *Government Information Quarterly* and *Annual International Conference on Digital*

*Government Research* or the interdisciplinary journals *Nature, Science,* and *Philosophical Transactions of the Royal Society.* These 5 outlets were chosen because they are considered to be leading research journals in terms of prestige and interdisciplinarity; the time period 2018 to 2023 meanwhile was chosen pragmatically, as it aligns with our interest in contemporary theoretical trends in the literature and because prior systematic reviews of AI in government (e.g., Sousa et al., 2019) exclusively cover the time period 2000 to 2018, which marked the beginning of advances in modern AI (Norvig & Russell, 2020). We then used the keywords from these seed articles combined with the Boolean operator 'OR' as the list of search terms (see Fig. 1B) for the final literature search on GS. This search term selection procedure was adopted to ensure that we used all key terms synonymous with or related to 'artificial intelligence' and 'government' commonly used in influential articles.

To identify relevant papers, the search terms were used for screening an article's title and abstract. The search was limited to original publications published in the English language during the period 1 January 2018 and 11 May 2023. The first 1,000 articles were then retrieved, sorted, and the highest impact articles were selected, where impact relates to the number of citations. Specifically, of these 1,000 articles, 5 duplicates were removed, leaving an initial corpus of 995 articles from which articles belonging to the first quartile were selected (i.e., top 25%). Aside from the semantic unfeasibility of integrating all possible terms that have been used to study AI-GOV, this sampling strategy is based on the finding that the most widely referenced concepts in a research field tend to stem from or are referenced in the most influential articles (Bornmann et al., 2020). The specific threshold was in turn chosen to strike a balance between oversampling and retrieving a diverse range of initial search results whilst also generating a manageable sample in line with prior integrative reviews (e.g., Tsvetkova et al., 2017). After excluding an additional 30 articles that were books, we manually filtered out additional articles published in journals belonging to the biological and physical sciences and only kept those published in social science, humanities, technical (i.e., computer science) or interdisciplinary journals. This resulted in a sample of 92 articles; when factoring in the seed articles, the final number of articles included in this review is thus 99 articles published in 76 different journals. The dataset of articles, extracted concepts and all code to reproduce the results are available in an online repository (see Availability of data and materials statement).

### 3.2 Concept extraction and co-occurrence analysis

For the purposes of our study design, we defined a concept as an adjective or abstract noun that denotes an idea or a quality which can be taken to offer a point of view for understanding AI-based systems in government and their desired attributes (e.g., transparency). After constructing our final literature corpus and downloading each article, we extracted key concepts by first relying on automatic extraction to detect 'candidate keywords' (henceforth also 'concepts'), i.e., the most important words in each text, and then using manual review for validation (Fig. 1A). We adopted this automatic procedure as a first step to minimize the bias that comes with manual review and to increase the reliability and reproducibility of our research design. Specifically, we implemented the light-weight unsupervised automatic keyword extraction algorithm YAKE! (Campos et al., 2020) in Python, which rests on statistical text features extracted from single documents to select the most relevant keywords of a text. This unsupervised extraction method was chosen as it builds upon local text statistical features extracted from each document (i.e., article PDF); that is, it is domain-independent and does not require any training corpus. We set the parameter number of *n*-grams to 1 and maximum number of keywords per document to 10 (after experimenting with higher values and finding no difference in our results); default settings were kept for all other the parameters. To validate the results of this unsupervised approach and increase the internal validity of our results, we then manually reviewed the list of extracted keywords per article, excluding those terms that did not meet our definition of a concept, and read the abstract of each article to ensure no key concepts were missed.

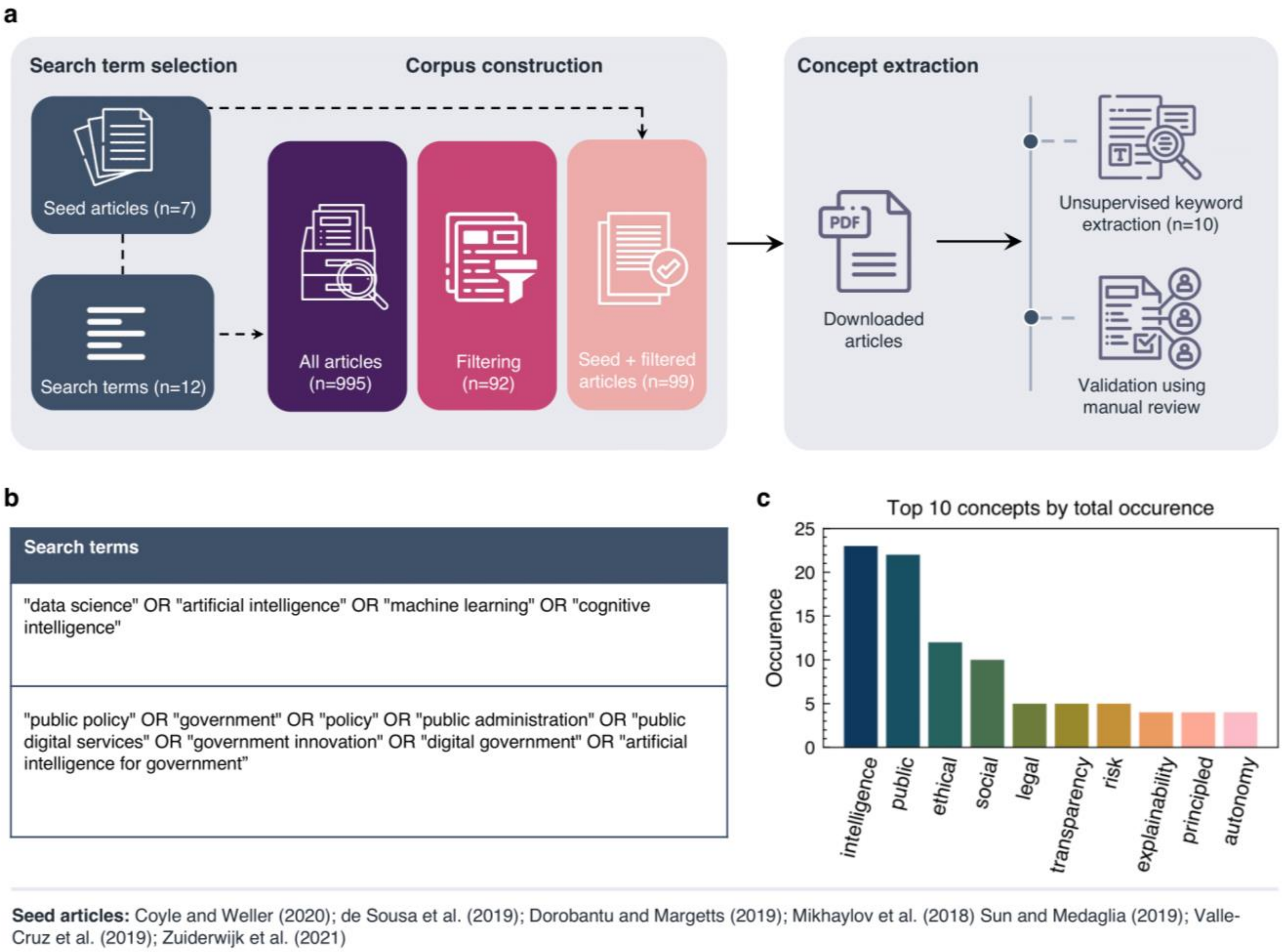

**Figure 1.** Literature search and concept extraction procedure. **(a)** Search term selection involved hand selecting seed articles (n=7) on the topic of AI in government and using the keywords (n=12) from these articles as search terms for a literature search. Filtering out articles that did not meet our inclusion criteria resulted in a final corpus of 99 articles from which key concepts were extracted using the unsupervised keyword extraction combined with manual abstract review. **(b)** The 12 search terms used for the literature search. **(b)** The top 10 concepts that we extracted. Each bar represents the total number of occurrences of a term in the final list of author-validated extracted concepts.

To understand not only which concepts frequently feature in the study of AI in government but also how they relate to each other, we then conducted keyword co-occurrence analysis (Newman, 2018). As a method that is extensively used to study the intellectual structure across fields (e.g., Tang et al., 2017), a co-occurrence network aims to represent the cumulative knowledge of a domain and helps to uncover meaningful insights based on the patterns of links between keywords 25/10/2023 19:34:00. This is based on the assumption that keywords are adequate descriptors of an article's content and can illustrate the ways certain ideas cluster together visually (Callon et al., 1991). The process involved computing the frequencies of co-occurrences between the author-validated concepts, constructing a co-occurrence matrix and visualizing the network to find central concepts and clusters. We followed K. Liu et al. (2020) by first performing vocabulary control to normalize spelling variations and synonyms of the same concept and applied a minimum concept co-occurrence threshold of 1 to include all potential co-occurrences (i.e., in Fig. 2 we only exclude concepts that do not co-occur with any other terms). The network analysis software Gephi (Bastian et al., 2009) was in turn used for visualizing the results, where each concept constitutes a network node and the edge weight is determined by their association strength, which normalizes the strength of pairwise co-occurrence by the frequency of each term (Eck & Waltman, 2009). Finally, the network analysis modularity-based community detection method (Blondel et al., 2008) was applied to sort concepts into distinct clusters. The overall network can in turn be said to provide a macro level understanding of AI in government scholarship, while each cluster represents distinct research themes, topics for which frequently co-occurring keywords have a shared bibliometric and semantic relationship (Cambrosio et al., 1993).

# 4 Results

In total, automatic extraction resulted in 990 candidate keywords, out of which 69 terms from 87 articles in 69 journals (51 social science and humanities journals, 14 technical venues and 4 interdisciplinary outlets) met our conception definition inclusion criteria. The journals with the most extracted concepts are *Government Information Quarterly* (number of extracted concepts=5) and *Business Horizons* (n=5) followed by *Nature* (n=3) and *Science* (n=3). Each one of the final list of concepts are taken to be relevant to a public sector context, pertinent now, or set trends for future work on AI-GOV. A list of the top 10 most frequently appearing concepts is shown in Fig. 1C; Table 1 provides definitions of each. Each definition seeks to capture how the respective term is used in the articles that we find and in turn stems either from an extracted journal article or, in the case where a journal article does not define the given term, a standard reference source. Choosing only concepts that co-occur at least once, the final co-occurrence network, shown in Figure 2, consists of a total of 64 nodes (concepts) and 120 edges (relations), meaning 5 concepts do not co-occur with any other concepts. The average degree of nodes in the network (i.e., number of relations that are incident on that node) is 3.75.

## 4.1 Overall trends

As shown in Fig. 2, four keywords act as bridging terms in the keyword cooccurrence network, connecting the clusters and network together to a high degree. These are: 'intelligence' (number of relations=23), 'public' (number of relations=22), 'ethical' (number of relations=12), and 'social' (number of relations=10). As the first two are explicitly included in the search terms and frequently co-occur with the words 'artificial' and government-related terms like 'administration' or 'policy', respectively, this result is perhaps to be expected. Similarly, the relative frequency of ethical and social, alongside the concept 'legal' (which occurs 5 times), likely stems from the fact that these three terms are often grouped together in articles on governing or regulating AI (e.g., Cath, 2018). Moreover, these terms predominate in discussions of AI in social science and humanities articles (Sloane & Moss, 2019), which make up the majority of our sample in terms of journal scope. Yet, what is perhaps most noteworthy are the terms that make up the remaining list of the top 10 most frequently occurring terms (see Fig. 1C); these are ‘transparency’

**Table 1.** Descriptions of the top 10 most frequently occurring concepts used in scholarship on AI in government based on the results of our integrative literature search and concept extraction procedure.

| Concept | Description |
|---|---|
| Intelligence | “[The] ability to interpret external data correctly, to learn from such data, and to use those learnings to achieve specific goals and tasks through flexible adaptation” (Haenlein and Kaplan 2019) |
| Public | “Of or relating to the people as a whole; that belongs to, affects, or concerns the community or the nation” (Oxford University Press, 2023) |
| Ethical | “Ethical AI is an AI that performs and behaves ethically ... [abiding by] moral principles governing the behaviours or actions of an individual or a group of individual” (Siau and Wang 2020) |
| Social | “[Relating to society,] what is best for us, what is best for others around us and what is best for society as a whole” (Perc et al. 2019) |
| Legal | “[Relating to] legal-regulatory governance solutions [for AI systems] … for example around liability regarding automated vehicles” (Cath 2018) |
| Transparency | “The interpretability of a given AI system, i.e, the ability to know how and why a model performed the way it did in a specific context and therefore to understand the rationale behind its decision or behaviour” (Leslie 2019) |
| Explainability | “[Associated with the notion of explanation as an interface between humans and a decision maker, explainability has] multiple meanings that can range from causal accounts and post hoc interpretations of decision to assurance that outcomes are reliable or fair in terms of the specified objectives for the system” (Coyle and Weller, 2020) |
| Principled | “[Associated with there being] embedded normative considerations in (AI) technology design and governance” (Mittelstadt 2019) |
| Autonomy | “[Possessing the ability to achieve a] goal within a defined scope without human interventions while adapting to operational and environmental conditions” (NIST in Ezenkwu & Starkey, 2019) |

Where possible, definitions are sourced from an article from which the respective term has been extracted. Words in square brackets are authors’ own additions to enhance definitional clarity.

and ‘risk’, which each occur 5 times; along with ‘explainability’, ‘principled’ and ‘autonomy’, each occurring 3 times. As indicated by the articles from which these concepts were extracted (see, e.g., Coyle & Weller, 2020 for a discussion of explainability), these terms are used across social and technical disciplines, and all have featured in high-level regulatory statements on how to promote AI (e.g., OECD, 2022).

Despite finding that 64 (93%) of the concepts we extract co-occur with at least 1 one other, the co-occurrence network can overall be characterised as sparse, as it has much fewer edges than the possible maximum number of edges (Goswami et al., 2016). This is supported by the fact that 5 concepts do not co-occur with any others (and hence are not included in Fig. 2): ‘judgment’, ‘executive’, ‘expertise’, ‘environmental’, and ‘renewable’. Given the semantic similarity of the first three of these terms, which all broadly relate to decision-making and knowledge reasoning, it may be speculated that the infrequent occurrence of these terms in contemporary scholarship mirrors the dominance of (and terminology associated with) data-driven, ML approaches to AI, over work on symbolic AI (Buchanan, 1990), which explicitly focuses on formal reasoning and building experts systems. In a similar vein, the infrequent occurrence and lack of a relation between the terms environmental and renewable to all other concepts is perhaps reflective of the fact that the literature on sustainable AI is still nascent. That is, while there is a growing effort towards understanding AI *for* sustainability (e.g., towards the sustainable development goals) (Vinuesa et al., 2020), research on the environmental impacts of actual AI systems is arguably still lacking (van Wynsberghe, 2021). One famous study (Strubell et al., 2019), for instance, showed that the process of training a single, deep learning, natural language processing (NLP) model (GPU) can lead to approx. 600,000 lb of carbon dioxide emissions.

**Figure 2.** Concept co-occurrence network graph of scholarship on AI in government. Each concept constitutes a node and the edge weight is determined by their association strength, which normalizes the strength of pairwise co-occurrence by the frequency of each tern. Concepts are grouped into 5 main clusters.

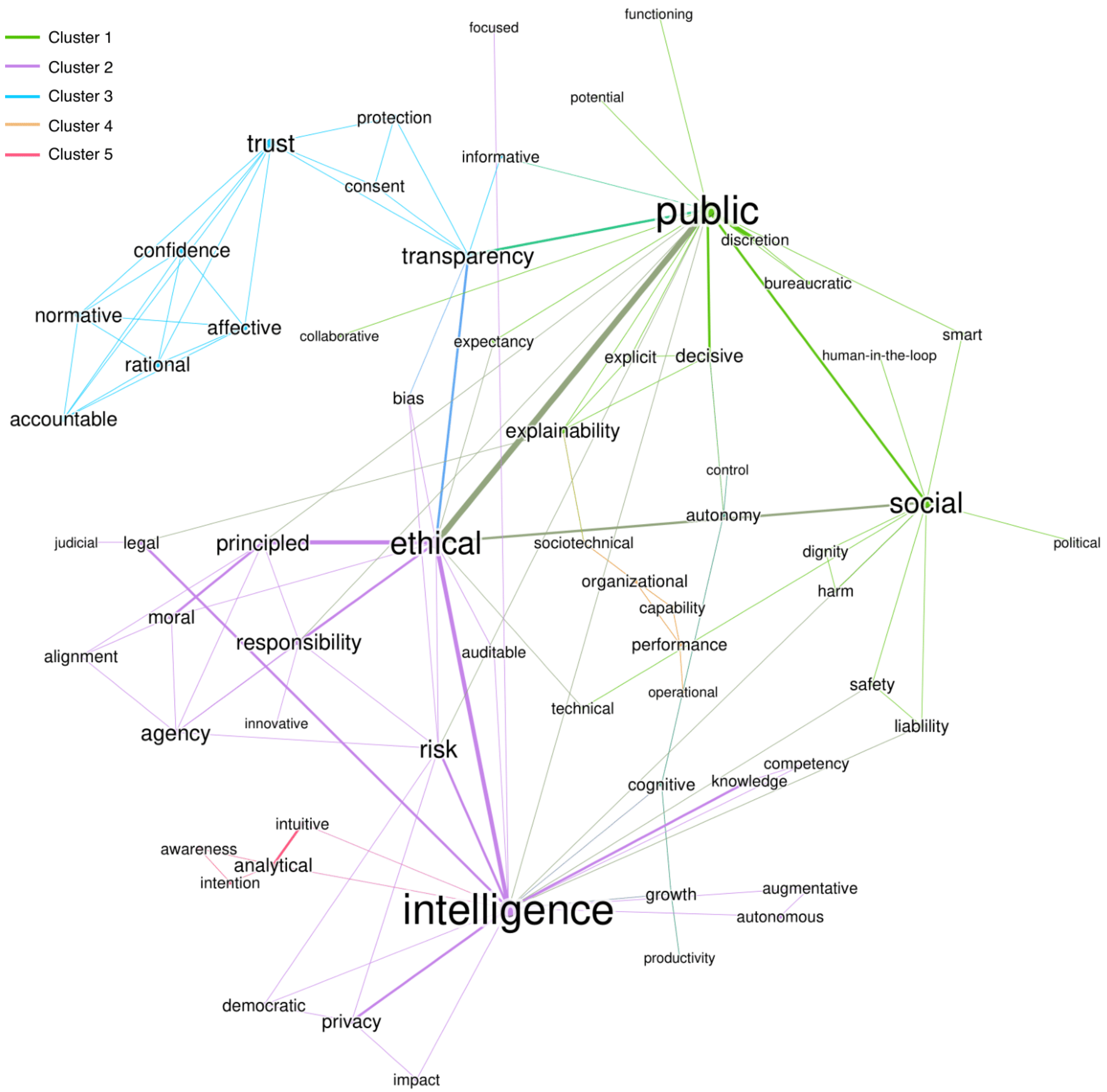

## 4.2 Concept co-occurrence clusters

Turning to the concept co-occurrence clustering results, modularity analysis of the co-occurrence network produced 5 main clusters of concepts (domains), assumed herein to represent distinct subject areas. A distinction can further be made between the three bigger green, purple and blue clusters (see Fig. 2 for colour reference) that include 86% of all concepts, and the two smaller orange and red clusters that include the remaining 14%. Of the 64 concepts extracted from the corpus of articles that co-occurred with at least 1 other concept, 23 (36%) were grouped into cluster 1 (green), in which two of the bridging terms, 'public' and 'social', appear the most. While 22 (34%) were grouped into cluster 2 (purple), in which the other two bridging terms 'intelligence' and 'ethical' predominate, and 10 (16%) are contained in cluster 3 (blue), in which 'trust' occurs the most, with a total of 3 occurrences. Finally, cluster 4 (orange) and cluster 5 (red) contain 5 (8%) and 4 (6%) concepts, respectively. The concept 'analytical' occurs the most in cluster 4 with a total of 3 occurrences, while the concepts 'organizational', 'performance', and 'sociotechnical' in cluster 5 each occur 2 times.

While the contents of the clusters are in part contingent on specific design choices made as part of our literature search and the scope of our review, the clusters seem to reveal distinct cohesive subject areas that exist within contemporary scholarship on AI in government. Cluster 1 (green), which may be labelled "societal and public expectations", itself contains two subclusters around the terms public and social that include a variety of concepts that all broadly relate to decision-making processes, such as 'discretion', 'decisive', 'control', 'bureaucratic', 'human-in-the-loop', 'explainability', and terms that can be said to denote societal and social justice connotations like 'harm', 'safety', 'political', and 'dignity', respectively. The primary disciplines that these concepts appear to be featured in most, as reflected by the source articles of the extracted concepts in terms of journal and author affiliations, are public administration, human-computer interaction, and ethics of AI. This perhaps reflects the fact that while concepts related to the social and ethical implications of AI have tended to be the purview of the social sciences and humanities (Miller, 2019), more recently, the subfields of AI, ML, and robotics have all contributed to their use and development. This can be seen by the fact that concepts such as explainability and safety have become thriving areas of research within AI in their own right (Laufer et al., 2022).

Similar to cluster 1, cluster 2 (purple), also contains two subclusters around the two most frequently occurring terms, ethical and intelligence, hence we label this cluster "ethical implications of intelligence". The other concepts connected most closely to the term ethical are terms relating to moral principles and practices including 'principled', 'responsibility', and 'bias', alongside 'moral', 'auditable', and 'alignment'. Within the subcluster around the term intelligence meanwhile, we find terms related to knowledge and behavioural attributes of AI-GOV including 'knowledge', 'competency', and 'autonomous', alongside less domain-specific terms including 'risk', 'privacy', and 'democratic'. Taken together, the fact that the term ethical and intelligence co-occur potentially indicates that, as AI systems become more advanced, scholarship on the latest advancements in the capabilities of AI systems increasingly also feature scholarly discussions on the broader and long-term ethical implications of these new capabilities (Clarke & Whittlestone, 2022).

In cluster 3 (blue), which we label "epistemic validation concerns", the term 'transparency' occurs the most, with a total of 5 occurrences, followed by 'trust', which occurs 3 times. Other concepts that can be found in this cluster, include 'confidence', 'rational', 'accountable', 'normative', and 'affective' (also see Fig. 2), all of which appear to be densely connected (i.e., co-occur together). Each of these concepts can be said to broadly relate to how AI-GOV are perceived and understood. Specifically, they seem to describe properties and values that pertain to the interface between AI applications and human actors. Both in terms of the knowledge, beliefs, and intentions of those using AI applications (e.g., a desire for trust), and the properties of the system itself (e.g., how transparent its architecture is.) Given that AI-GOV represent a step-change from ICTs due, among other factors, to their increased technical complexity, these concepts appear to have increased in usage and importance as this complexity has grown.

Of the final two smaller clusters, cluster 4 (orange), contains the more general descriptive terms 'organizational', 'performance', 'capability', 'sociotechnical', and 'operational'. Whilst cluster 5 (red) includes the concepts 'analytical', 'awareness', 'intuitive', and 'intention', which all seem to describe specific aspects of intelligence. Yet, the small size of both clusters, makes it harder to consider both as cohesive research domains within the field of AI in government; this is substantiated by the fact that the articles from which concepts in cluster 5 were extracted constitute only 2% of the corpus of all articles.

Overall, these results indicate that there are a myriad of influential concepts and a number of distinct clusters in current scholarship on AI in government, with contributions from multiple fields spanning social and technical disciplines. However, despite the multidisciplinary nature of current research, we also find that concepts are sparsely connected. It is surprising, for example, that 'functioning' is not connected to 'intelligence' and 'trust' is not connected to 'public', as both of these pairs of terms are typically used in similar discussions. This lack of connections suggests there are still deficiencies in the literature, as scholars are not drawing all the necessary conceptual links between concepts. Nevertheless, given the purposively narrow focus of our research design, it is also important to reiterate that our mapping is not exhaustive hence multiple concepts that are less well-referenced or not included within our corpus of influential articles are not included. 'Fairness', 'controllability', and 'interpretability' are a notable examples to name but a few (for a comprehensive recent glossary that provides defections of these, and other omitted terms, see Estevez Almenzar et al. 2022). As such, the field will likely benefit from additional, systemic reviews to identify all links between terms.

## 5 Towards a unified conception of AI-GOV

Based on our mapping of literature on AI in government, we contend that a new conceptual framework is needed to integrate the diverse conceptual strands that currently permeate scholarship. In the rest of this paper, we try to so in the following way: we first propose three new, multi-faceted concepts for studying AI-GOV, before using these to construct a novel conceptual typology of AI-GOV. A 'typology', sometimes called a 'property space', is defined both as a method to classify observations in terms of their attributes, and as the development of theories about configurations of variables that constitute conceptual types. We understand and use the term primarily in the latter sense: as an organized system that breaks down an overarching phenomenon—in this case AI-GOV —into component dimensions (George & Bennet, 2005). To develop our typology, we follow O'Raghallaigh (et al., 2010), and focus on exploring the dimensionality and refining the measurement of the concepts we propose, as opposed to exhaustively classifying all different types of AI-GOV. Moreover, in proposing multi-faceted concepts, constituted of multiple elements, we take inspiration from the recommendations from a recent review of how prior studies have attempted to classify AI systems, which stresses the need to integrate information and offer concepts that can be expected to remain stable and useful over time (Mökander et al., 2022).

It is important to acknowledge that many of the frequently occurring key terms within current scholarship (see Table 1) are themselves essentially contested and problematic;

explainability, for example, has multiple meanings and is often used interchangeably with the term ‘interpretability’, despite subtle semantic differences in meaning (see Molnar, 2020 for a discussion). Similarly, ‘public’ is a very broad term that will likely be used within AI-GOV literature in multiple ways. Yet, while new concepts are regularly introduced to resolve such ambiguity issues, many still approach the subject from the perspective of a single discipline, such as ethics in the case of ‘explicability’ (Floridi & Cowls, 2022), or law in the case of ‘artificial immutability’ (Wachter, 2022). In contrast, we work towards bridging past efforts to conceptualize AI-GOV by abstracting away the essential characteristics of the terms we map, before combining the new concepts we introduce into a unified typological framework. Our system of classification in turn aims to establish an informative and productive connection between the dimensions used and situates each within their semantic field. Hence, from here on out we also refer to the concepts as dimensions. The labels we give to the three new concepts we propose are: (1) *operational fitness*, (2) *epistemic alignment*, and (3) *normative divergence*. In Section 5.1 below we first provide definitions for each concept, Section 5.2 then introduces our proposed conceptual typology of AI-GOV. Finally, Section 5.3 proceeds to discuss the practical implications of our typology by drawing connections to emerging technical measurement standards development by Standards Development Organisations (SDOs) and other organisations around world, while also considering the limitations and challenges of operationalization.

## 5.1 Three new multifaceted dimensions for the study of AI-GOV

Theorizing in public administration has largely been dominated by single concept, ideal type constructs (Bullock et al., 2022). However, based on the results of our concept mapping and co-occurrence analysis, we argue that the multidisciplinary nature of AI-GOV calls for the development of new, multifaceted concepts that better capture the complexities of the sociotechnical phenomena of AI-GOV and encourage cross-disciplinary dialogue. Our conceptualisation is thus motivated by the wish to reach a multidisciplinary audience and the desire to abide by recommended social science criteria for concept development. For achieving the latter, we have, therefore, followed Gerring (1999), and ensured our concepts are bounded (i.e., operationalizable), possess depth (i.e., accompany properties shared by the instances under definition), and offer theoretical utility (i.e., are useful within a wider field of inferences).

### *Operational fitness*

The first dimension we propose is *operational fitness*. It focuses on the functional abilities and potential performance advantages an AI system might possess over existing systems, in cognitive, computational, and organisational terms. More specifically, it relates to the internal computational capabilities of a system, the extent to which the system can be expected to operate in accordance with standards of routine functioning, and how well it performs at a given cognitive task when compared to other agents. Formally, we define it as:

| Dimension | Definition |
|---|---|
| Operational fitness | The degree to which the composition and functions of a government system incorporating an AI application (or set of applications) aligns with (1) codified standards of organisation, system construction and functioning required to operate in a particular environment, and the extent to which it can be expected to (2) outperform other agents at a specific cognitive task (or set of tasks). |
| Epistemic alignment | The degree to which (1) data and information about the composition of a government system incorporating an AI application (or set of applications) align with standards of knowledge sharing, and (2) its behaviour in every circumstance is known by all affected parties. |
| Normative divergence | The degree to which the observed behaviour of a government system incorporating an AI application (or set of applications) in its current environment, including the consequences of the system's behaviour, diverges from (1) formal institutional standards and (2) affected parties' perceptions of acceptable behaviour. |
| Each dimension can be applied to study systems at two scales of inquiry: those incorporating a single AI application, and those incorporating multiple AI applications. Affected parties of an AI-based system are likely to include developers (designers, engineers, and domain experts), managers (public administrators and street-level bureaucrats), and end-users affected by the AI-based system at a systemic, societal level (citizens). | |

**Table 2.** Proposed dimensions to analyse and classify AI systems in government.

> *The degree to which the composition and functions of a government system incorporating an AI application (or set of applications) aligns with (1) codified standards of organisation and system construction and functioning required to operate in a particular environment, and the extent to which it can be expected to (2) outperform other agents at a specific cognitive task (or set of tasks)*

As AI systems approach or exceed "human-level" accuracy in an increasing number of domains (Bengio et al., 2021; Nguyen et al., 2018; Shinde & Shah, 2018), reliable claims about the relative success rates of AI systems will depend on rigorous assessments of their behaviour and performance (Cowley et al., 2022). Depending on the task, this may involve developing new methods for comparing humans and AI-based systems in "species-fair" comparisons (Firestone, 2020), which would account for the differing performance constraints faced by competing agents, or developing new measurement benchmarks when using AI and data science for government tasks that humans cannot do (MacArthur et al., 2022). Operational fitness aims to take this into account by considering both the competence (internal capabilities) of a system and how it performs in comparisons with others.

Similar to some of the terms featured in cluster 1 ("ethical implications of intelligence") as well as clusters 4 and 5 (e.g., performance and awareness), operational fitness aims to provide a proximate view of whether the system in its current form is functionally fit for purpose. In management terms, it aims to describe how well a system operates in the face of both external and internal constraints. Following Simon (1995), an analogy can also be drawn here between measuring the performance of an AI system and the biological concept of 'Darwinian fitness', in that nature generates systems or modifies existing ones then tests their ability to survive in the ambient environment (ibid). AI-GOV experience the same process, except that the generator (the design process) is more purposeful, and the tests (the purposes the designers have in mind) go beyond biological fitness. In sum, operational fitness, therefore, aims to describe whether a particular AI-GOV, based simply on its design, can be expected to effectively function, and continue operating as intended.

### *Epistemic alignment*

The second dimension we propose, which we call *epistemic alignment*, relates to how affected parties, and human actors in general, perceive and know the functions of a particular AI-GOV functions. It responds to the observation that our ability to use and solve problems using AI to make progress depends heavily on epistemic processes: how information is produced and distributed, and the tools and processes we use to make decisions and evaluate claims (Clarke & Whittlestone, 2022). Formally, we define it as:

> *The degree to which (1) data and information about the composition of a government system incorporating an AI application (or set of applications) align with standards of knowledge sharing, and (2) its behaviour in every circumstance is known by all affected parties.*

A system's level of epistemic alignment thereby relates to the key cluster 3 concepts introduced in Section 3, such as transparency and trust. Once operationalized, it aims to provide evidence as to whether claims about the operational fitness of a system are warranted. That is, in describing affected parties' knowledge of a particular AI-GOV, it can be used to gauge both whether the operational behaviour of a system is understood, and how much is known. Affected parties are likely to include system developers, those who manage and operate them within government, and end-users affected by the AI system. Importantly, codified norms and standard of knowledge-sharing may themselves be incomplete, as such, a system can be in full alignment with a particular standard of transparency but that standard itself may be deficient or outdated and need to be updated over time. In other words, the degree to which an AI system is epistemically aligned depends on how well knowledge-related standards themselves are defined (this is somewhat analogous to the challenge of 'outer alignment' in AI alignment research, that is, effectively specifying the desired objective a system should seek to achieve).

***Normative divergence***

The third and final dimension we propose, which we label *normative divergence*, is meant to reflect the need to establish whether an AI system operates in accordance with evaluated standards of acceptable behaviour, as determined by the affected parties of an AI system. As such, it relates most to concepts grouped together in cluster 1 ("societal and public expectations"). Because AI-GOV applications are ultimately still used and judged by humans, any conceptualization of AI-GOV needs to consider, at least at a theoretical level, the difficulty affected parties will have in determining ethical standards and governance methods for how an application incorporating an AI system will be used, overseen, and monitored. Normative divergence, which aims to do just that, is in turn formally defined as:

> *The degree to which the observed behaviour of a government system incorporating an AI application (or set of applications) in its current environment, including the consequences of the system's behaviour, diverges from (1) formal institutional standards and (2) affected parties' perceptions of acceptable behaviour.*

Once operationalized and measured, an AI-based system's level of normative divergence thereby aims to provide evidence as to whether the system in its current form is deemed acceptable and can be expected to persist. Importantly, the normative divergence of a particular AI-GOV may, in

**Table 3.** Characteristics of proposed dimensions to classify AI-based systems in government.

| Dimension | Dimension levels | Description | Example measurement standards |
|---|---|---|---|
| Operational fitness | Basic | The AI-based system meets critical benchmarks in core operational fitness standards, especially in terms of effectiveness. Application components satisfy core objectives | • ISO/IEC DTS 4213.2 Assessment of Machine Learning Classification Performance<br>• Singapore AI Governance Testing Framework Toolkit<br>• ISO/IEC TR 24029-1:2021 Artificial Intelligence (AI) — Assessment of the robustness of neural networks<br>• ISO/IEC TR 24027:2021 Information technology — Artificial intelligence (AI) — Bias in AI systems and AI aided decision making<br>• Model cards for model reporting in Mitchell et al., 2019 |
| | Intermediate | The AI-based system exceeds critical benchmarks in core operational fitness standards, especially in terms of effectiveness and safety. Application components satisfy core objectives | |
| | Advanced | The AI-based system meets or exceeds benchmarks in all operational fitness standards, including in terms of effectiveness, safety, reliability, usability, and adaptability. Application components satisfy all objectives | |
| Epistemic alignment | Nominal | Data and information about the AI-based system meets critical benchmarks in core epistemic alignment standards, especially in terms of transparency. Knowledge about the system is incomplete | • UK Algorithmic Transparency Standard<br>• UK Metadata Standards for Sharing and Publishing Data<br>• IEEE Standard for XAI – eXplainable Artificial Intelligence - for Achieving Clarity and Interoperability of AI Systems Design*<br>• IEEE 7001-2021 Standard for Transparency of Autonomous Systems<br>• Datasheets in (Mitchell et al., 2019)** |
| | Dispersed | Data and information about the AI-based system exceeds critical benchmarks in core epistemic alignment standards, especially in terms of transparency and explainability. Knowledge about the system is distributed unequally among affected parties | |
| | Uniform | Data and information about the AI-based system meets or exceeds benchmarks in all epistemic alignment standards, including in terms of transparency, explainability, reproducibility, interpretability, and openness. All relevant knowledge about the system is distributed appropriately among affected parties | |
| Normative divergence | Low | The behaviour of the AI-based system consistently converges with codified standards and affected parties' perceptions of acceptable behaviour | • Algorithm Charter for Aotearoa New Zealand<br>• UK Public Attitudes to Data and AI Tracker Survey<br>• EU AI Alliance<br>• Canadian Algorithmic Impact Assessment Tool<br>• ISO/IEC 38507:2022 Information technology — Governance of IT — Governance implications of the use of artificial intelligence by organizations<br>• ForHumanity Independent Audit of AI Systems (IAAIS) |
| | Moderate | The behaviour of the AI-based system consistently converges with codified standards of acceptable behaviour, but there is variance in affected parties' perceptions | |
| | High | The behaviour of the AI-based system consistently diverges from codified standards of acceptable behaviour. Operational fitness or epistemic alignment may need to be re-evaluated. If normative divergence remains universal, other changes may have to be made | |

All example measurement standards reflect naming conventions at the time of writing. *Under development. **Emerging industry standard.

many cases, only be determined locally, as the exact affected parties' norms and values can vary greatly between applications contexts (i.e., administrative regions) (Ehret, 2022).

## 5.2 A conceptual typology of government AI-based systems

Our typology of government AI-based systems is formed of three dimensions, introduced above: operational fitness, epistemic alignment, and normative variance. In making use of these dimensions, our typology serves two main functions. It is both theory-building in the sense that it serves to characterize AI-GOV and further explicate the meaning of each dimension, and classificatory, in that it places examples of different types of AI-GOV in conceptually appropriate cells. While the latter function is secondary and hence examples of AI-GOV are primarily illustrative, when combined and expanded, we contend that these three principal dimensions can be used to encapsulate the entire hypothetical space of AI-GOV and effectively discriminate different types of system. Moreover, by drawing links between our proposed scales and actual measurement standards, the typology aims to aid academics, policymakers, and engineers in developing an understanding of the full space of AI-GOV and think through the application of a specific AI system in a unified manner, taking into account technical, legal, and ethical concerns. Table 3 provides full-length definitions of the scales proposed for each dimension and relates each to emerging measurements standards; Figure 3 presents a visualization of the typology with example types of AI-GOV for illustration.

To construct our typology, we scale each of our dimensions. For conceptual parsimony, we adopt a 3-point scale for each dimension. We adopt this rule, as opposed to a 2-point (i.e., 'high' vs 'low') scale typically found in many social science typologies, to better capture the wide range of potential AI-GOV, when conceptualized using the dimensions we propose. We divide operational fitness into *basic, intermediate,* and *advanced* where a particular AI-GOV is classified as possessing basic operational fitness if *it meets critical benchmarks in core operational fitness standards and application components satisfy core objectives*. These core standards and objectives are expected to include, at a minimum, effectiveness, which can be broadly defined as "the extent to which [a system] achieves its intended objectives" (Salamon, 2002). We focus on effectiveness as we consider it is a prerequisite for all the other aspects of operational fitness. Consequently, a system is understood to exhibit advanced operational fitness when it *meets or exceeds benchmarks in all operational fitness standards, including in terms of effectiveness, safety, reliability, usability, and adaptability.*

As for epistemic alignment, we distinguish between AI-GOV that can be characterised as possessing either *nominal*, *dispersed*, or *uniform* epistemic alignment. Similar to operational fitness, nominal epistemic alignment here *means data and information about the AI-based system meets critical benchmarks in core epistemic alignment standards, especially in terms of transparency,* which is taken to be a prerequisite for other epistemic domain aspects of a system, and *knowledge about the system is incomplete*. In more concrete terms, information about a particular AI-GOV may be transparent but not fully reproducible or explainable. In contrast, a system can be considered to exhibit uniform epistemic alignment when *data and information about the AI-based system meets or exceeds benchmarks in all epistemic alignment standards, including in terms of transparency, explainability, reproducibility, interpretability, and openness, and all relevant knowledge about the system is distributed appropriately among affected parties.* In the latter sense, an analogy can be drawn between uniform epistemic alignment and the concept of 'perfect information' in microeconomics, specifically game theory, which describes the situation when all agents have complete and instantaneous knowledge of all market prices, their own utility, etc. (Fudenberg & Tirole, 1991). However, the important difference is that, given uniform epistemic alignment indicates a state of affairs where each party (agent) affected by a particular AI system has the appropriate knowledge it needs to interact with and use that system, this allows for different parties may have different levels of knowledge.

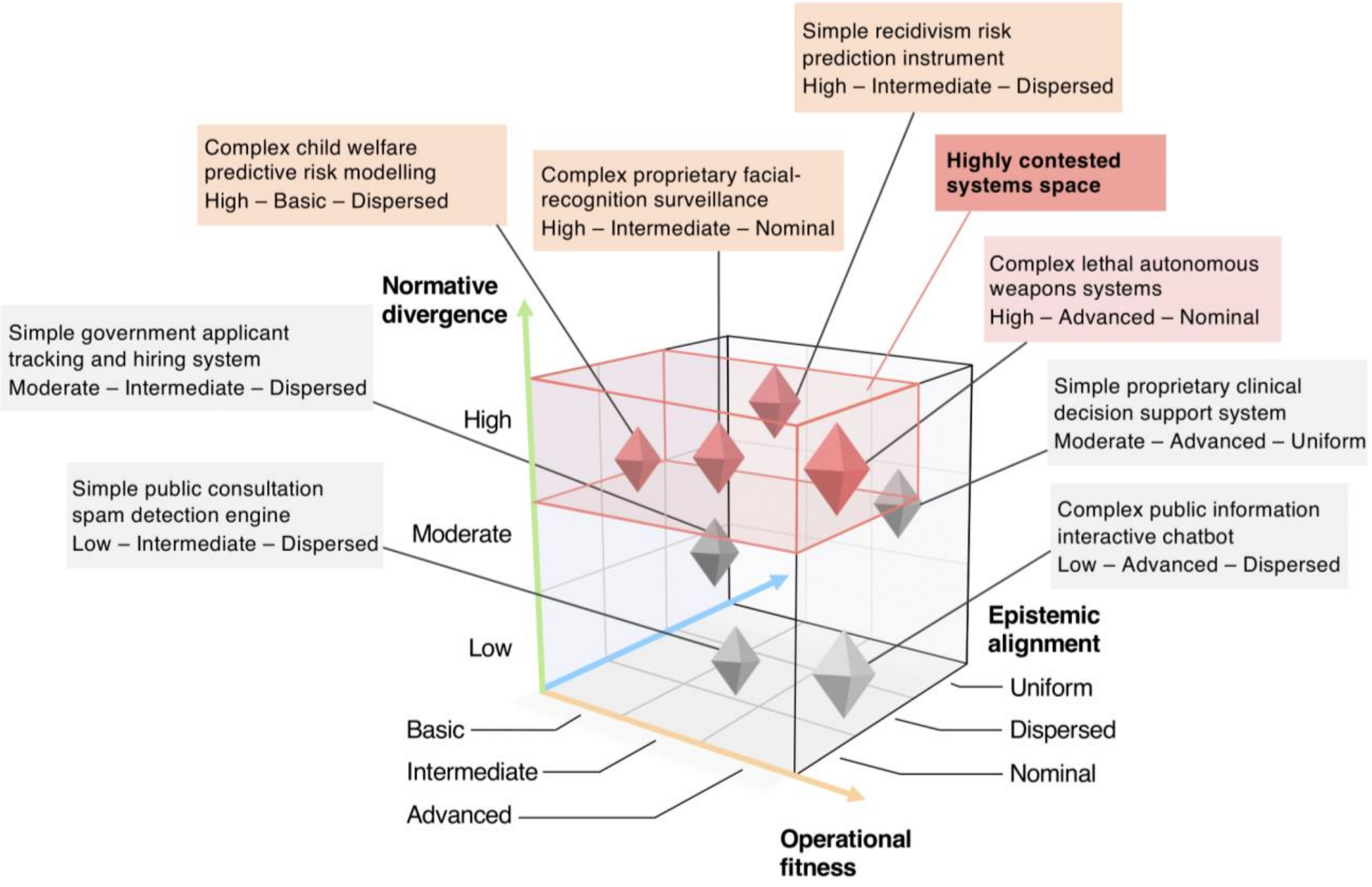

**Figure 3.** Classifying types of AI-based systems in government. A classification schema places select types of AI-based systems in government along three main axes: normative divergence, operational fitness, and epistemic alignment. Systems incorporating complex ML models (e.g., neural networks, boosted decision trees, random forests) are contrasted with those using simpler ones (e.g., logistic regression, decision lists). 'Complex lethal autonomous weapons systems' are classed as highly contested systems (red box), which exhibit high normative divergence and possess advanced operational fitness. A 3-point scale for each dimension is adopted for conceptual parsimony; it is anticipated that future standards for grading real-world systems may use different scales. The EU AI Act, for instance, uses a 4-point scale to classify the risk level of an AI system. Where possible, placement of systems is informed by surveys of public attitudes towards AI (e.g., Nussberger et al., 2022), otherwise systems are placed using authors' own interpretation.

Normative divergence meanwhile is separated so that AI-GOV can be differentiated into *low*, *moderate*, or *high* systems where a system is considered to enjoy low normative divergence if *the behaviour of the AI-based system is consistently in alignment with codified standards and affected parties' perceptions of acceptable behaviour*. Given the at times highly localized nature of moral viewpoints about AI and decisions made by machines, key normative standards and affected parties' perceptions will likely depend on the context in which a particular AI-GOV is expected to operate. Notably, at least when it comes to moral dilemmas faced by autonomous vehicles, past studies on cross-cultural variation in ethical and moral preferences have nevertheless uncovered major clusters of countries that may in fact share many normative standards (Awad et al., 2018). In line with this finding, a system is considered to possess high normative divergence when *the behaviour of the AI-based system is inconsistently in alignment with codified standards of acceptable behaviour*. In practice, this may be the case when a system is found to break near universal standards of human rights and freedoms, such as Universal Declaration of Human Rights (UN General Assembly, 1948), which have already informed emerging legal frameworks for the design, development, and deployment of AI systems more generally (Leslie et al., 2021).

Having explicated the scales of the dimensions that make up our typology, we can now consider what combining each to capture the conceptual space of AI-GOV may look like. For analytical simplicity and clarity, we only consider systems that can be deployed in practice. In other words, the different types of AI-GOV our typology maps out are all assumed to possess basic operational fitness. For the sake of conceptual clarity, we assume that AI-GOV applications that 'fail' to achieve this basic criterion are retracted and hence cannot be considered as part of the space of possible systems. While there are many factors that influence whether a system is cancelled, this assumption is broadly in line with a recent comparative study of why government departments decide to pause or cancel their use of automated decision systems (Redden et al., 2022); the authors find that half of the decisions to pause or cancel the use of systems relate to concerns about the lack of effectiveness of a system.

Figure 3 places select types of AI-GOV in the typology space as defined by the three main axes, operational fitness, epistemic alignment, and normative divergence. For illustration, we differentiate, among others, between those systems that have advanced operational fitness, nominal epistemic alignment, and high normative divergence, signified by the example of complex lethal autonomous weapons systems, and those which have advanced, dispersed, and low values for each of these dimensions, respectively. An example of the former may be systems that incorporate armed drones that can search for and eliminate enemy combatants in a city. The latter may include complex public information recommendation engines, such as the GOV.UK Related Links systems adopted by the UK government to aid navigation of its online services (UK Government, 2022). Both can be understood to be complex in so far as the AI system may consist of components (i.e., an ML model) that are not interpretable or easily reproducible (e.g., neural networks, boosted decision trees, or random forests). They differ, however, in terms of their exact operational fitness (e.g., effectiveness, accuracy, etc.) and normative divergence (e.g., political feasibility, non-maleficence etc.). While only a small number of illustrative types are depicted in Figure 3, there are by now countless examples of AI-GOV that we contend can all be placed into the conceptual schema it shows (Engstrom et al., 2020).

## 5.3 Practical implications: The need for global standards

While the purpose of our typology is primarily conceptual, a growing number of AI measurement standards make the possibility of operationalizing the dimensions we propose to compare different AI-GOV currently in use increasingly plausible. Specifically, if the dimensions are taken to be composite, consisting of multiple measurement scales, then we need only consider the various standards that are emerging for some of the concepts mapped in Section 4. Focusing on operational fitness, for instance, two emerging instruments are helping to create a global standard of how to measure AI systems in terms of robustness: the ISO/IEC DTS 4213.2 and the ISO/IEC TR 24029-1:2021 standards [*] (see Table 3). When combined with other measurement standards for performance it may soon be possible to compare AI-GOV in terms of whether they exhibit basic, intermediate, or advanced operational fitness. Similarly, for epistemic alignment, the UK Algorithmic Transparency Standard, and the IEEE Standard for XAI – eXplainable Artificial Intelligence are aiming to provide universally-applicable methods to measure transparency and explainability, respectively. Finally, even though normative divergence is arguably the hardest dimension to standardize, given norms and perceptions of AI-GOV can rapidly change and are often highly contextual, several initiatives are emerging to decide how particular systems may be graded in terms of normative concern around trustworthiness, sustainability, and fairness, among other key concepts. For instance, the Algorithm Charter for Aotearoa New Zealand involves a clear, formal, and ongoing 'affected parties engagement strategy' including with public agencies likely to use the policy and urges designers of AI-GOV to consult those impacted by their use (Ada Lovelace Institute et al., 2021).

[*] The full titles are 'ISO/IEC DTS 4213.2—Assessment of Machine Learning Classification Performance' and 'ISO/IEC TR 24029-1:2021 Artificial Intelligence (AI)—Assessment of the robustness of neural networks', respectively.

Fully operationalizing and standardizing each of the dimensions we propose hinges on both the development of further global standards alongside those developed in specific locales and government contexts. This is especially relevant for measuring normative divergence, which is intrinsically connected to the application context. Developers, regulators, and commissioners need to understand the nature and implications of AI-GOV currently being developed and assess the adequacy of regulatory standards. Promisingly, current legal attempts, such as the draft European Union AI Act, the first legal framework on AI (Sovrano et al., 2022), aim to set new obligations to ensure coherency across many of the key concepts that make up each of the dimensions proposed herein, such as risk and fairness. Elsewhere, efforts such as the UK's Plan for Digital Regulation show how regulators are aiming to meet the challenge of regulating activities transformed by AI (Aitken et al., 2022). While ongoing and often context-neutral, this standardization process shows the possibility of linking each of the dimensions in a composite manner to accepted standards so that a systems operational fitness or epistemic fitness may be approximated.

Importantly, it has to be acknowledged that it may not always be possible to operationalize normative divergence for all contexts, as some AI systems may require attributes (e.g., accountability) that are only inherently meaningful when attached to human operators, given that the machine components of AI can arguably not be granted the status of legal agents (Brożek & Jakubiec, 2017). Although we place governance and accountability structures and standards within the dimension of normative divergence within our typology, we thus recognise the significant variation and levels of specification that conceptions of such processes have in use. By providing a multi-layered approach to governance and highlighting that governance of AI-GOV needs to be viewed and assessed through three different dimensions, our typology can thus also highlight inconsistencies or process conflicts when it comes to regulating real-world use cases if AI.

AI-GOV are necessarily highly contextualised and nested within institutional contexts of use with internal standards, bounded departmental or regulatory remits and established processes. As a result, global standards of governance and accountability may provide an initial macro-level developmental base from which to operationalise and problematise risk which are then expanded to incorporate the specific meso-level institutional or contextual governance processes. Alternatively, existing micro-level oversight and governance assessment processes may be developed to incorporate relevant global standards for AI. An example of such development work is the algorithmic impact assessment for data access in a healthcare context developed by the Ada Lovelace Institute (Ada Lovelace Institute, et al., 2021). This design proposal for algorithmic impact assessments is focused upon primarily normative divergence considerations within a high-stakes context; however, as the report outlines it sits alongside established standards and regulations which cover areas including data protection, ISO clinical standards for assessment of clinical performance and medical device regulation which span all of the typology dimensions outlined.

Taken together, by outlining the entire application space and specifying the minimal conditions that must be met for an AI system to be deemed suitable for government use, our typological framework aims to be useful in guiding operational, epistemic, and normative theorizing about, and development of, AI-GOV, including future standards for each of these three domains. Only through the use of standards can AI applications be embedded into government using clear operational procedures and epistemic criteria and behave in alignment with the normative expectations of society. The typology further aims to stress the need for multidisciplinary analysis in the study of contemporary applications of data-dependent AI, and highlights that there is no single 'ideal' type of AI system that can handle every task; instead, there is always a complex interplay between a system's operational fitness, epistemic alignment, and normative divergence. Through promulgating this line of reasoning, the typology can in turn inform official language used to describe AI and government deliberation on AI adoption in publication administration in a more unified way and encourage practitioners to approach issues around human oversight and the adaptative behaviour of AI-GOV in a more informed manner. AI-GOV are not static but entities that can change and evolve over time. It is, therefore, crucial that governments continuously assess the behaviour of AI-GOV and minimize their potential to cause harm. With the rapid development and adoption of current AI technologies, especially generative AI, the urgency of this cannot be understated; a recent survey of 1,600 researchers finds, for instance, that scientists are concerned, as well as excited, by the increasing use of AI. At the same time, surveys of public attitudes indicate that the public find various potential use cases, such as autonomous weapons, to be very concerning and want immediate regulation of AI (Ada Lovelace Institute and The Alan Turing Institute, 2023).

## 6 Conclusion

This paper set out from the premise that recent advances in AI, especially in generative language modelling, hold the promise of transforming government administration and public service delivery, given that state-of-the-art systems are able to execute an increasing range of government-related occupational tasks. Likely impacts on the citizenry that governments service in turn include a greater demand for personalised services as well as the increased use of forecasting within policy-making to meet local and individual needs, to name but two (Margetts and Dorobantu 2019). Yet, when taking a global view, the use of AI in government is still in its early phase of implementation. Therefore, nascent decisions about how to conceptualize AI-GOV will greatly shape government expectations. When designed and implemented correctly, such tools could help deliver services and set policies with unprecedented speed and computing abilities. At the same time, the adoption of AI may exacerbate biases and precipitate harms. The responsible use of AI requires clarity about the characteristics of the systems built around them. By proposing the multifaceted concepts of operational fitness, epistemic alignment, and normative divergence alongside a conceptual typology to analyse AI, this article aims to stimulate informed debate among scholars and practitioners aiming to rethink public administration with AI. To conclude, we highlight some of the implications and limitations of our typology and highlight future research avenues that we anticipate will become more pressing in future.

One key path forward for the field of AI in government in general to is to gather more empirical work to enrich the concepts we map and propose. It is especially important to complement critical reflections and conceptual analyses with empirical research in the field of public administration to gain a better understanding of how AI affects governmental routines, decisions, and bureaucratic agency. In this respect, the concepts and typology of AI-GOV we present invites future research to consider the relation between the various dimensions. For instance, will systems that exhibit advanced operational fitness and high normative divergence necessarily remain contested, or can this tension be resolved?

Another way to consolidate the study AI in government is by moving towards more global technical standards of AI, which can benefit further conceptual work and vice versa. The ultimate goal for AI in government is a healthy ecosystem of theoretical work, empirical case studies, sandboxes of AI in government, and cutting-edge use-cases which all feedback into each other. Notably, technical standards are rapidly already becoming key instruments for both the development and adoption, as wells the governance of AI technologies. One field where a lot of progress is being made, and which can inspire AI in government, is medicine and healthcare. Here, an internationally developed set of standards for assessing the quality of clinical trials involving AI already exists (Cruz Rivera et al., 2020; Liu et al., 2020) and the use of regulations, standards and practice for the market in medical devices is well-established (Ramakrishna et al., 2015). As discussed in Section 5, standards relating to normative divergence are especially scarce and governments would greatly benefit from taking a more structured approach to understanding the concerns and viewpoints of diverse affected parties before adopting an AI system. In a UK context, one recent development is the launch of the AI Standards Hub, an initiative that is part of the UK's National AI Strategy and which is dedicated to the evolving and international field of standardisation for AI technologies. The hub includes a Standards Database[†] that covers nearly 300 relevant standards that are being developed or have been published by a range of prominent SDOs.

The use of AI in the delivery of government services is an ongoing development which will expand in the next few years. As stated at the outset, it is critical that these developments are embedded using standard operational procedures, clear epistemic criteria, and are always in line with normative concerns. To help realise this aim, we need to analyse AI-GOV from multiple levels. Our work offers an initial contribution toward achieving this goal by providing a richer conceptualization of AI in government that we hope can spark further work.

[†] Available at: https://aistandardshub.org/ai-standards-search/.

## Availability of data and materials

Code and data used as part of the analysis and needed to reproduce the initial search results can be accessed through the project folder at the open science framework (OSF) site: https://osf.io/8z4kh/. Raw data can also be accessed by using the search terms provided in the methods section and repeating the search in Google Scholar using the code in the OSF project site, which is also hosted in the following GitHub repository along with further documentation: https://github.com/ai-for-public-services/ai-gov-framework.